\documentstyle[prb,aps,preprint]{revtex}
\begin{document}
\draft
\title{Ultrasonic evidence of an uncorrelated cluster formation temperature
in manganites with first-order magnetic transition at $T_C$}
\author{J. Mira and J. Rivas}
\address{Departamento de
F\'\i{}sica Aplicada, Universidade de Santiago de Compostela, E-15782 Santiago
de Compostela, Spain}
\author{A. Moreno-Gobbi and M. P\'erez Macho\footnote{On leave from
Departamento de F\'\i{}sica Aplicada, Universidade de Santiago de Compostela,
E-15782 Santiago de Compostela, Spain}}
\address{Departamento de F\'\i{}sica de los Materiales  -
Facultad de Ciencias, Igu\'a 4225, 11400 Montevideo, Uruguay}
\author{F. Rivadulla}
\address{Texas Materials Institute, ETC 9.102,
The University of Texas at Austin, Austin, Texas 78712, USA}
\maketitle

\begin{abstract}

Ultrasonic attenuation and phase velocity measurements have been carried out in
the ferromagnetic perovskites La$_{2/3}$Ca$_{1/3}$MnO$_3$ and
La$_{2/3}$Sr$_{1/3}$MnO$_3$. Data show that the transition at the Curie
temperature, $T_C$, changes from first- to second-order as Sr replaces Ca
in the perovskite. The compound with first-order transition shows also
another transition at a temperature $T^* > T_C$. We interpret the temperature
window $T_C < T < T^*$ as a region of coexistence of a phase separated regime of
metallic and insulating regions, in the line of recent theoretical proposals.

\end{abstract}


\section{Introduction}

The advances in the comprehension of the physical properties of manganese
perovskites have been given in two main boosts. The first one started in 1950,
when these materials were first considered by Jonker and van Santen.
\cite{Jonker} In that period, that spanned over more than a decade, the basic
mechanisms governing their magnetic and transport properties were defined,
giving
rise to the description of the double exchange
interaction.\cite{Zener,Anderson,deGennes} The second one took place last
decade, after the discovery of colossal magnetoresistance
(CMR)\cite{VH,Jin} (although a high MR had already been observed experimentally
before, first by Searle and Wang\cite{Searle} and later by Kusters {\it et
al.}\cite{Kusters}). During this period, in which we are still immersed, the
research activity reached a magnitude that made the topic a starring one in the
community of Condensed Matter Physics, and the current opinion about the
state-of-the-art underwent several changes. In the first few years after the
description of CMR, explanations were tried on the basis of double
exchange,\cite{Furukawa} but it was demostrated soon that such effect alone
could not explain the magnitude of CMR.\cite{Millis} Later, certain evidences
in
favor of the existence of an inhomogeneous electronic state in these
materials started to
appear.\cite
{Uehara,Fath,Merithew,Raquet,Podzorov,Kapusta,Kim,Yunoki,Moreo}
This tendence continues nowadays after the conclusions of many groups that
found
-and are still finding- that the homogeneous description of the magnetic and
electronic state of manganites  is almost ruled out.

In order to establish the
basis for such behavior the idea of phase separation, i.e., the coexistence of
regions of localized and itinerant carriers within the same
crystallographic
structure, has been proposed.\cite{GoodenoughNW,MoreoScience} In this
framework,
Dagotto, Hotta and Moreo\cite{reviewDagotto} proposed in a recent review the
existence of a temperature window where coexisting metallic and
insulating clusters appear above the Curie temperature, $T_C$, when this
transition is first-order and there is a source of disorder. This temperature
window would extend up to $T^* > T_C$, where these clusters
dissociate to form two-Mn Zener polarons (or small polarons). The aim of the
present work is to find experimental evidence of such temperature $T^*$. For this
purpose, ultrasonic attenuation and phase velocity measurements were carried out
in La$_{2/3}$Ca$_{1/3}$MnO$_3$ and La$_{2/3}$Sr$_{1/3}$MnO$_3$. This technique is
proved to be a very sensitive tool not only for studying defects and
microscopic
processes in solids, but also for probing systems undergoing magnetic and
structural phase transitions.\cite{Garland}

\section{Experimental details}

Samples were prepared by solid state reaction of La$_2$O$_3$,
CaO or SrCO$_3$, MnO$_2$ and MnO (at least 99.995 $\%$ in purity), which were
heated in air in two steps (1100$^\circ$C 70 hours, 1200$^\circ$C 27 hours) and
pressed into disks. The temperature was slowly ramped at 5$^\circ$C/min, and
cooled down to room temperature at 2$^\circ$C/min. Intermediate grindings were
made. Pellets were finally pressed at 7 Ton/cm$^2$ and annealed at 1300$^\circ$C
for 100 hours, with an
intermediate grinding at 30 hours. The nominal oxygen content was almost
stoichiometric as determined by iodometric analysis. X-ray
powder patterns were collected at room temperature using a Philips PW1710
diffractometer, working with $CuK \alpha$ radiation. The lattice parameters,
derived by Rietveld analysis, are in agreement with those reported in the
literature.

Cylindrical pellets, with
a diameter of 10.0 mm and heights 2.4 mm (LaCaMnO), and 3.1 mm (LaSrMnO) were
used to perform ultrasonic measurements. The two opposite faces of each
sample were polished so that the difference between two points in each surface
was not more than 1 $\mu$m. The ultrasonic velocity and attenuation
measurements were performed on a Matec-6600 series by means of the conventional
pulse-echo technique. Y-cut quartz transducers of 5 MHz fundamental frequency
were used for transverse ultrasonic excitation. They were coupled to the sample
surface with a grease for high temperature uses (leak point 550 K). The sample
was coupled at high temperature, and then cooled until 200 K and maintained at
this temperature for 15 minutes before measurementes were carried out. The
ultrasonic elapsed time for a pulse round trip (transit time) was obtained with
the pulse-echo-overlap technique at the initial temperature. Then, time
variation with temperature was automatically monitored with a sensitivity of
0.05 ns at 5 MHz. The experiment was taken in a closed-cycle refrigerator
(Janis) and the temperature of the sample was varied at a rate of about 0.5
K/min.

\section{Results and discussion}

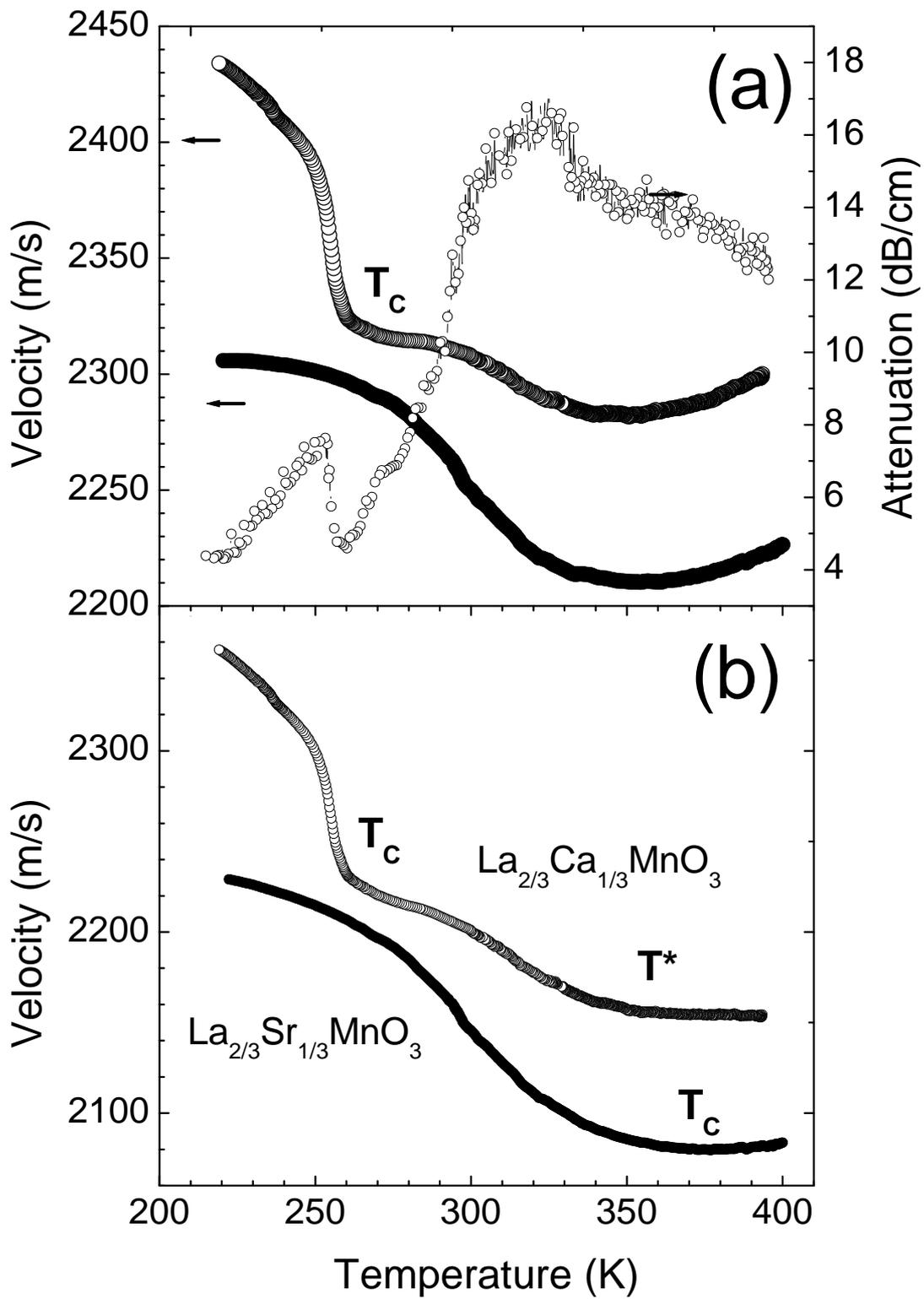
\begin{figure}  
\caption
{(a) Transversal ultrasonic velocity vs. temperature of
La$_{2/3}$Ca$_{1/3}$MnO$_3$ (open symbols) and La$_{2/3}$Sr$_{1/3}$MnO$_3$
(filled symbols), and attenuation vs. temperature of
La$_{2/3}$Ca$_{1/3}$MnO$_3$ (attenuation not measurable in LaSrMnO), as measured.
 (b) Ultrasonic velocity vs. temperature for the
same samples after background substraction. Note the
steep decrease of velocity as well as the peak in attenuation at the T$_C$ of
LaCaMnO. Note also the similar velocity anomalies in the region 350-370 K,
corresponding to $T^*$ in LaCaMnO and to T$_C$ in LaSrMnO.}                
\label{un}     
\end{figure}

In Fig. \ref{un}(a) we show ultrasonic velocity vs. temperature data for
La$_{2/3}$Ca$_{1/3}$MnO$_3$ and La$_{2/3}$Sr$_{1/3}$MnO$_3$, and attenuation
vs. temperature data for La$_{2/3}$Ca$_{1/3}$MnO$_3$, obtained with 5 MHz
transversal waves between 220 and 400 K. For LaSrMnO it was impossible to
measure the attenuation because its high value allowed to obtain only two echoes
at room temperature (at higher temperatures the second echo became negligible).
The transit time was measurable at room temperature, and then time variations
were measured with our automated improved Matec system. The influence of thermal
expansion has not been taken into account, because it is  negligible: if
($\delta$$l$)/l=$2 \cdot 10^{-4}$ we have that for our sample, with $l$= 2.4 mm
thick (4.8 mm microwave path) and a velocity of about 2500 m/s, $\delta t$
($t$=time) is $5\cdot10^{-4}$ s and therefore, at T$_C$,  the
associated variation of velocity is far smaller than the one we measured.

In our original ultrasonic velocity data, an approximately linear
increment with temperature was observed at temperatures above 350 K for both
samples. At this temperature range no mechanism related with magnetic phase
transitions is expected, and this variation should be attributed to another
mechanism, as the produced by crystalline defects present in the
material.\cite{Nowick} The resonant mechanism associated to dislocation lines is
well known from crystalline metals. As Moreno-Gobbi and
Eiras\cite{Ariel}, working on crystalline copper, have shown, it
gives a term proportional to T for velocity and proportional to T$^2$ for
attenuation. The coefficient of proportionality depends on $L^2$ for
velocity and $L^4$ for attenuation, where $L$ is the average free length of
the segment of dislocation lines. In the present paper we assume a background of
this kind in order to take into account the variation of velocity at high
temperatures. The velocity background was fitted to the high temperature region
of velocity vs. temperature curves, and then substracted from original data.
These fits gave a term $0.37$ T for LaCaMnO and $0.46$ T for LaSrMnO. The
so-corrected velocity data are presented in Fig. \ref{un}(b). The differences
between the numerical values of the coefficients of proportionality have their
origins in the different values of $L$ expected for different
materials.\cite{Nowick}

The ultrasonic velocity for La$_{2/3}$Ca$_{1/3}$MnO$_3$ presents two clear
anomalies: an abrupt hardening at about 250-260 K, and another lesser at about
350-360 K. The first one marks the T$_C$ of this
system\cite{Hwang,Lynn,Savosta2,Mira1,Mira2} (T$_C$= 260 K). The behavior is
similar to that found at MHz frequencies by other authors,\cite{Ramirez,Zhu} and
at kHz frequencies by Cordero {\it et al}.\cite{Cordero} The abrupt decay of the
curve at T$_C$ signals the first-order character of this transition,\cite{Juffe}
an interpretation that is in agreement with results obtained from other
techniques.\cite{Hwang,Lynn,Savosta2,Mira1,Mira2} The decay in velocity is
accompanied by a peak in the attenuation curve at T$_C$. This peak is
asymmetric, in agreement with the results of Cordero {\it et al.},\cite{Cordero}
who explain it invoking the presence of inhomogeneous phases below T$_C$ in a
similar way to relaxor ferroelectrics.\cite{Viehland,Lu} 

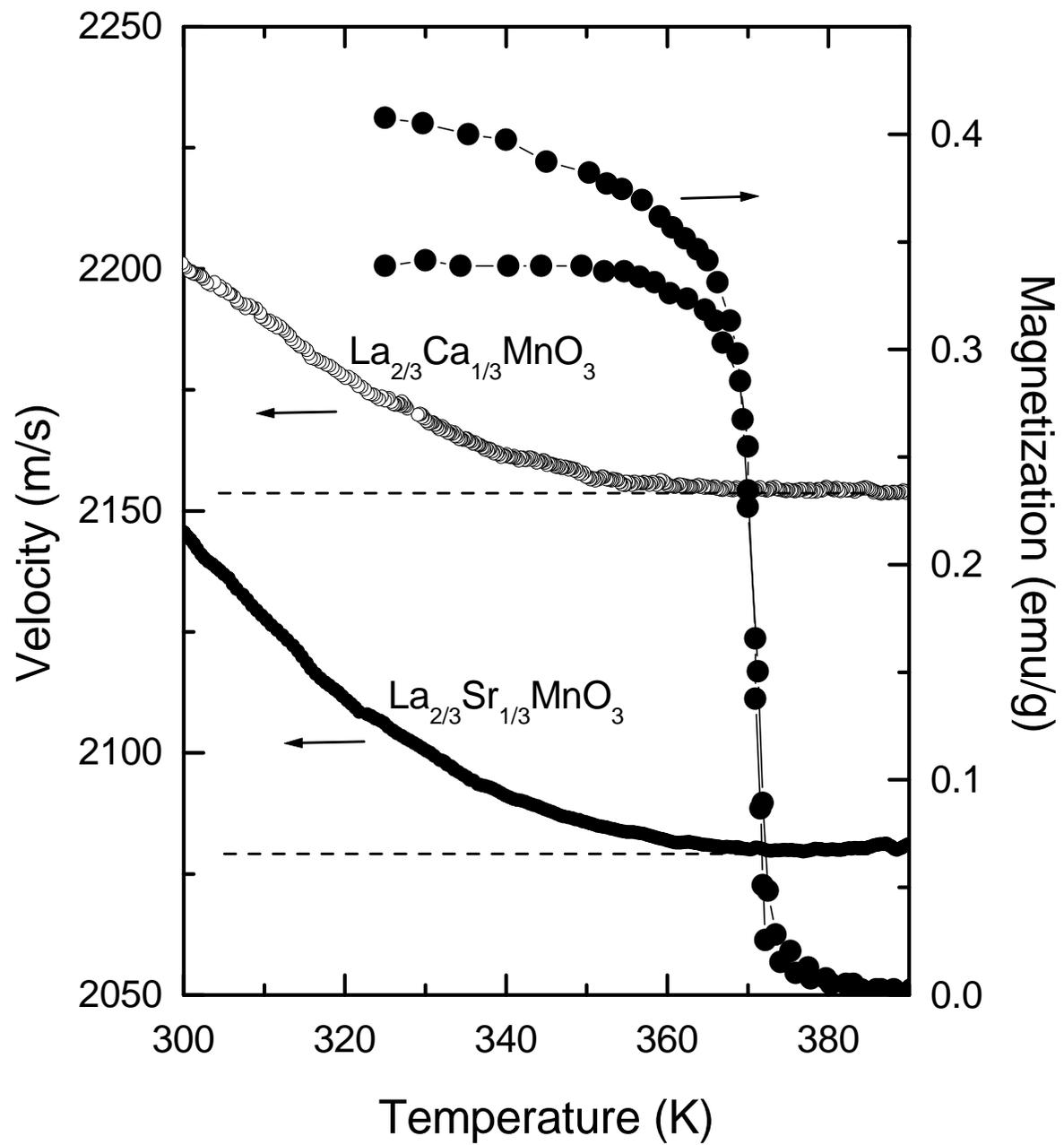
\begin{figure}  
\caption
{Detail of the corrected transversal ultrasonic velocity  vs. temperature of
La$_{2/3}$Ca$_{1/3}$MnO$_3$ and La$_{2/3}$Sr$_{1/3}$MnO$_3$ in the $T^*$ region.
For comparison, magnetization vs. temperature data (zero-field-cooled and
field-cooled) of La$_{2/3}$Sr$_{1/3}$MnO$_3$ are included.}               
\label{dous}      
\end{figure}

The second anomaly takes place above T$_C$, showing a hardening process at 
$T^*$ $\simeq$ 350-360 K (Fig. \ref{dous}). In Fig. \ref{un} it is also observed a
very large attenuation peak at 320 K accompanying this velocity anomaly. This
was not observed by Cordero {\it et al.} at kHz frequencies, but is clearly
observed in our MHz measurements. The reason lies on the higher sensitivity of
MHz techniques to small inhomogeneities, as magnetic clusters, due to the
shortest wavelength (approximately 0.5 mm at 5 MHz for transversal waves).

In Figs. \ref{un} and \ref{dous} it is also observed that the only noticeable
anomaly in the utrasonic velocity curve of La$_{2/3}$Sr$_{1/3}$MnO$_3$ is a
hardening process beginning at about 370 K (in the order of that of LaCaMnO in
the same temperature region), that is coincident with the second-order phase
transition point of this compound, clearly observed by magnetization
measurements (T$_C \simeq$
370 K).\cite{Mira1,Lofland,Lofland2,Martin,Mohan,Ghosh}  In the explored
temperatures, no more transitions were detected. Based on this behavior we can
perfectly associate the process of hardening observed at 370 K to this
second-order phase transition.

One of the open issues in the comprehension of the physics of ferromagnetic
manganese perovskites at present is to determine the nature of the magnetic
phase above $T_C$. From our ultrasonic velocity data it seems that this phase
is different in the two systems considered. First of all,
whereas LaCaMnO
displays a first-order transition at $T_C$, LaSrMnO presents a second-order
one. This difference, that has already been highlighted by magnetic
measurements,\cite{Mira1} and attributed by Mira {\it et al.}\cite{Mira2} to
a change in the crystal symmetry, from orthorhombic to rhombohedral, seems to
have further implications. Magnetoresistance, thermal expansion, magnetovolume
and magnetocaloric effects experience a considerable change when moving from
one compound to the other.\cite{Mira2,Ramos,Mira3} Also, calorimetric data
showed evidence for an anomaly at a similar $T^*$ in those
manganites
La$_{1-x}$(Ca,Sr)$_x$MnO$_3$ with a first-order transformation at
$T_C$,\cite{Mira2} in agreement with
the ultrasonic data presented here that show the existence of a transition point
at $T^*>T_C$ in LaCaMnO. It is worth mentioning that our $T^*$ is similar to the
point at which both the volume thermal expansion deviates from a Gruneisen fit
and the inverse magnetic susceptibility deviates from a Curie-Weiss fit, after
results of de Teresa {\it et al.} in La$_{2/3}$Ca$_{1/3}$MnO$_3$ (Ref.
\cite{deTeresa}). Supporting our results, we have to mention that neutron
diffraction studies\cite{Viret} in the paramagnetic region of
La$_{2/3}$Sr$_{1/3}$MnO$_3$ show differences with respect to
La$_{2/3}$Ca$_{1/3}$MnO$_3$. We want also to call the attention on the similar
values of the temperatures of the second-order transitions, $T^*$ $\simeq$
350-360 K for La$_{2/3}$Ca$_{1/3}$MnO$_3$, and $T_C$ $\simeq$ 370 K for
La$_{2/3}$Sr$_{1/3}$MnO$_3$. In some sense, it leads to think that the magnetic
interaction between the Mn atoms is of the same strength in both compounds
and only the occurrence of the first-order change in LaCaMnO is breaking the
long-range order at a temperature below $T^*$. After R\"{o}der, Zhang and
Bishop,\cite{Roder} this might be due to lattice effects, that would decrease
the T$_C$ associated to the DE coupling.

The anomaly at $T^*$ shows that the magnetic phase transition is not
from ferromagnetism to a purely paramagnetic state, but some sort of magnetic
structure is present. Such idea is in accordance with the theoretical results
of
Moreo {\it et al.},\cite{Moreo} who have proposed, after computer simulations,
that large coexisting metallic and insulating clusters of equal electronic
density are generated in manganese oxides with first-order magnetic
transitions. Kimura {\it et al.}\cite{Kimura} have qualified these compounds as
relaxor ferromagnets, with a relaxation from one phase to the other. The
fluctuations between both phases have been seen by Transmission Electron
Microscopy by Podzorov {\it et al.}.\cite{Podzorov2} In such a case,
after Kimura {\it et al.}\cite{Kimura} the transition would be diffuse,
like the one observed by us at $T^*$. Nowadays there is being growing
experimental work that leads to assume the electronic phase-separated nature of
such materials, and now it is well established that this phase separation is not
due to chemical inhomogeneities or to the existence of phases with a different
chemical composition.\cite{Mathur} Although this approach is quite recent in
manganese perovskites, it had already been proposed by Se\~nar\'\i{}s
Rodr\'\i{}guez and Goodenough\cite{Tona} to explain the magnetic and electrical
properties of Sr-doped cobalt perovskites, thinking of a percolative transition
at T$_C$.

In summary, we consider that our data confirm the existence, in
La$_{2/3}$Ca$_{1/3}$MnO$_3$, of the temperature window $T_C<T<T^*$
proposed by Dagotto, Moreo and Hotta\cite{reviewDagotto}, where probably the
coexistence of a phase separated regime of metallic and insulating regions
could
be taking place. Such window is not observed in La$_{2/3}$Sr$_{1/3}$MnO$_3$, a
material with a conventional second-order ferromagnetic to paramagnetic phase
transition at its Curie temperature.

\acknowledgements

The DGI of the Ministry of Science of Technology of Spain is acknowledged for
financial support under project FEDER MAT2001-3749. FR also wants to acknowledge
the Fulbright foundation and MCYT of Spain for financial support.

\end{document}